\documentclass{icrc29}
\usepackage{graphicx,amssymb,amsmath,times}
\begin{document}
\title[Neutrino-induced muons observed with MINOS]{Neutrino-induced muons observed with MINOS}
\author[A. Habig et al.] {A. Habig$^a$,
  for the MINOS Collaboration$^b$\\
        (a) Univ. of Minnesota Duluth Physics Dept., 10 University Dr.,
        Duluth, MN 55812, USA\\
        (b) http://www-numi.fnal.gov/collab/collab.ps
        }
\presenter{Presenter: A. Habig (ahabig@umn.edu), \  
usa-habig-A-abs2-he22-oral}

\maketitle

\begin{abstract}

The Main Injector Neutrino Oscillation Search (MINOS) experiment's Far
Detector has been operational since July 2003, taking cosmic ray and
atmospheric neutrino data from its location in the Soudan Mine
Underground Lab.  Numerous neutrino-induced muons have been observed.
The detector's magnetic field allows the first determination by a large
underground detector of muon charge and thus neutrino versus
anti-neutrino on an event by event basis.

\end{abstract}

\section{Introduction} 

The MINOS experiment uses two similar iron/scintillator calorimeters to
measure the properties of the NuMI neutrino beam over a long baseline,
with a goal of precisely measuring the neutrino flavor oscillations seen
in atmospheric neutrinos~\cite{SK-full}.  The 5400~ton Far Detector is
located 700~m (2070~mwe) deep in the Soudan Mine Underground Lab in
northern Minnesota~\cite{icrc-detector}.  The rock overburden
reduces the rate of cosmic ray muons reaching the Lab by a factor of
$10^5$, allowing the detection of atmospheric $\nu_\mu$ via
their charged-current production of muons tracked in the detector.
These interactions can occur in the detector itself~\cite{icrc-keith} or
in the rock surrounding it.  While such muons entering the detector from
above cannot be distinguished from the $\sim0.5$~Hz of cosmic ray muons
penetrating the overburden, cosmic rays cannot penetrate large 
depths of rock so selecting horizontal and upward-going muons ensures
that they are neutrino-induced.  While other experiments have made such a
measurement in the past~\cite{upmus}, the MINOS detector is
magnetized with a toroidal field of $\sim1.5$~T, allowing the
determination of momentum and charge on an event-by-event basis via the
curvature of the muon track.  Muon momentum provides information on the
energy of the parent neutrino, and muon charge tags the parent neutrino
as a neutrino or anti-neutrino, something which has only been previously
measured in bulk~\cite{icrc-sk}.

\section{The Data}

To separate those neutrino-induced muons coming from below (at a rate of
a bit more than one per week) from those cosmic ray muons coming from
above (at a rate of one every couple seconds), timing information along
the track is used.  Topologically, the two classes of events look
identical - a series of flashes of light (localized to several cm in
space by the granularity of the scintillator strips) creating a track
which crosses the detector.  The only difference between upward- and
downward-going events is the times the light was deposited in the
scintillator, giving the direction the particle was headed along the
track.  This is expressed in terms of a measurement of the muon's
velocity $\beta = v/c$.  For highly relativistic down-going muons $\beta
\simeq +1$.  Upward-going muons have a negative velocity in this
convention, $\beta \simeq -1$.

\begin{figure}[h]
  \begin{center}
    \includegraphics[width=0.8\textwidth]{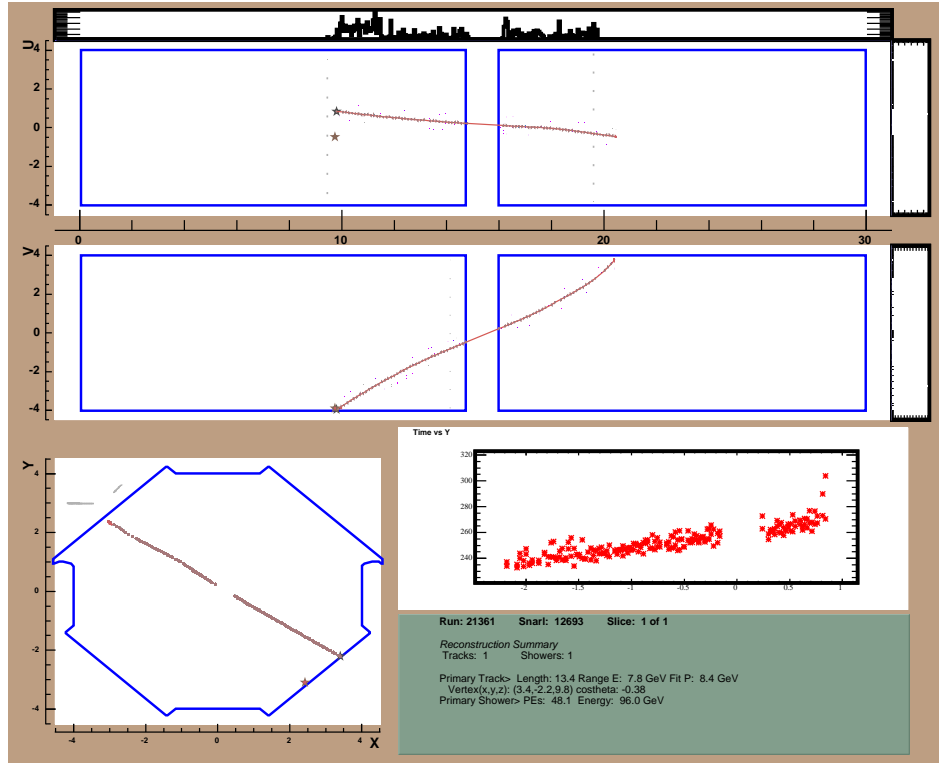}

    \caption{\label{fig:evd}
      An event display of an upward-going muon in the MINOS Far
      Detector.  The top two panels show the locations of digits along
      the length of the detector, in U and V coordinates with Z (along
      the length of the detector) increasing from left to right.  This
      event crosses the gap between the two halves of the detector.  The
      octagon in the lower left is the face-on view, in which this muon
      enters in the lower-right and exits from the upper left.  The
      graph in the lower right plots time vs detector ``Y''.  The time
      clearly increases with increasing height, showing that this event
      is moving up not down.  The histogram along the very top of the
      figure shows the scintillation light along the track.
      This muon track exhibits curvature, allowing the momentum
      (8.4~GeV/c) and sign ($\mu^+$) to be determined.
    }
  \end{center}
\end{figure}

The ability to tag each flash of scintillation light precisely in time
is crucial to this analysis.  The electronics used in MINOS have a
granularity of 1.56~ns and the rise time of a phototube signal is
$\sim2$~ns.  Signal propagation delays between different sections of
electronics can be tens of ns, but these timing offsets are easily
calculated using a clean sample of single cosmic ray muons, the geometry
of the detector, and the knowledge that these muons move close to the
speed of light.  The quality of the timing calibration is checked by
comparing the light arrival times at both ends of the same scintillator
strip with the true location of the hit obtained via tracking.  The
total uncertainty in the time for each hit or ``digit'' is
$\sigma_t=2.4$~ns.  An example of this direction determination can be
seen in the graph in the lower right of Fig.~\ref{fig:evd}.  This time
vs. height graph is clearly upward going.  The scatter of the points
about the trend is indicative of the timing uncertainty.

\begin{figure}[h]
  \begin{center}
    \includegraphics[width=0.8\textwidth]{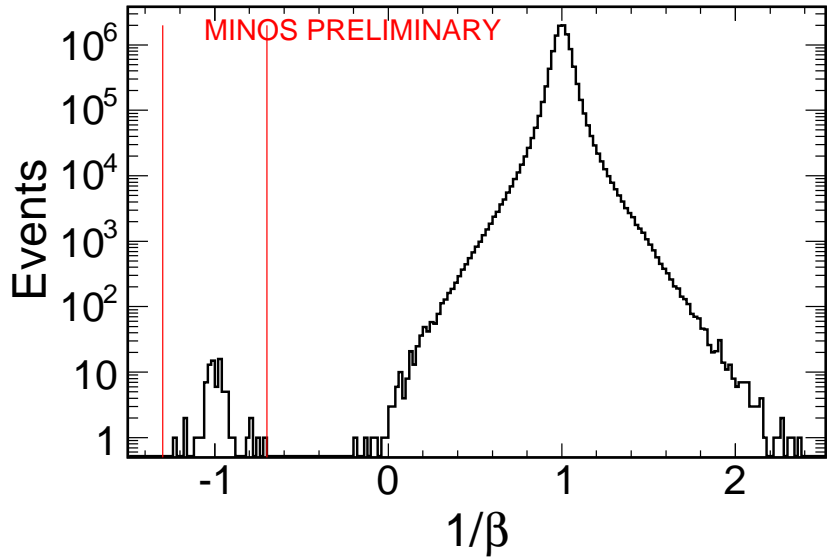}

    \caption{\label{fig:invbeta}
      The distribution of $1/\beta$ for the sample of crossing muon
      events seen in MINOS.  Downgoing cosmic ray muons have $\beta = +1
      \pm 0.049$ and up-going muons travel in the opposite direction
      with $\beta = -1 \pm 0.051$.  The vertical red lines are the cuts
      in $1/\beta$ used to select neutrino-induced upward-going muons.
    }

  \end{center}
\end{figure}

To ensure a well-determined data sample, the following cuts were applied
to the collection of muon tracks (for details see~\cite{brebel}).  A
single muon which crossed 20 planes over a length of at least 2.0~m must
have a well-determined spatial track fit, enter the detector, have a
well-determined time fit, and fall within the $1/\beta$ range shown in
Fig.~\ref{fig:invbeta}.  Data were collected from the completion of the
MINOS Far Detector in July 2003 through April 2005, spanning 464
live-days (160 of those days had a reversed magnetic field for
systematic checks).  91 upward-going muons were observed, a rate of 1
per 5.1 days.

\begin{figure}[h]
  \begin{center}

    \includegraphics[width=0.8\textwidth]{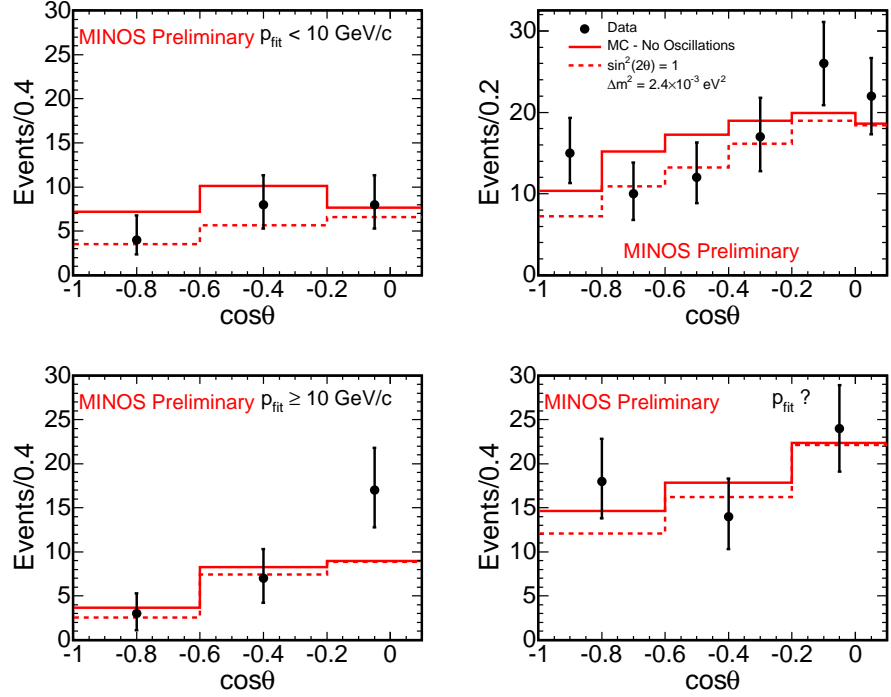}

    \caption{\label{fig:costheta}
      The distribution of neutrino-induced upward-going muon data as a
      function of zenith angle -- directly upward ($\cos\theta=-1$)
      through horizontal ($\cos\theta=0$) including some slightly
      downgoing (up to $\cos\theta=+0.1$). The data are the points, the
      no-oscillation prediction is the solid line, and the prediction of
      a $\sin^2 2\theta=1$, $\Delta m^2=2.4\times 10^{-3}~\rm{eV}^2$
      oscillation model are the dashed lines.  All muons are in the
      upper right plot, the subset for which no momentum was able to be
      determined the lower right, muons below 10~GeV/c on the upper
      left, and muons above 10~GeV/c on the lower left.
    }

  \end{center}
\end{figure}

The arrival direction of the muons is closely correlated with the
baseline of the parent neutrino, so data were plotted with respect to
the cosine of the zenith angle.  The momentum of the muon is correlated
to the energy of the parent neutrino, so the data were broken into three
samples, low ($<10$~GeV/c), high ($>10$~GeV/c), and indeterminate
momenta.  The last category tends to hold the highest momenta muons as
their tracks are too straight to clearly identify any curvature.  All the
data are in the upper right of Fig.~\ref{fig:costheta}, the other
plots separated by momenta.  Of the muons for
which a momenta was able to be determined, 25 were $\mu^-$ and 16 were
$\mu^+$.  

Monte Carlo (``MC'') data were created for comparison to the real data.
A 2500 year equivalent neutrino exposure was generated using the NUANCE
generator~\cite{nuance} with the Bartol96~\cite{bartol} input flux.  The
detector response has been carefully modeled with a GEANT~3 based
simulation, and the resulting MC data analyzed using the same
reconstruction tools and cuts as the real data.  Both the no-oscillation
hypothesis and the current Super-K best fit oscillation parameters are
also shown in Fig.~\ref{fig:costheta}.

\section{Conclusions}

A sample of neutrino-induced upward going muons has been collected by
the MINOS Far Detector.  91 events were observed over 464 live-days.
The detector's magnetic field allows the separation of these muons by
charge and by momentum.  The distribution of the muon arrival directions
in $\cos\theta$ was compared to Monte Carlo expectations.  The data and
expectations are consistent within the current statistical errors.

\section{Acknowledgments}

This work was supported by the U.S. Department of Energy, the U.K.
Particle Physics and Astronomy Research Council, and the State and
University of Minnesota.  We gratefully acknowledge the Minnesota
Department of Natural Resources for allowing us to use the facilities of
the Soudan Underground Mine State Park.  This presentation was directly
supported by NSF RUI grant \#0354848.

\end{document}